\documentclass[letterpaper]{article} % DO NOT CHANGE THIS
\usepackage{aaai24}  % DO NOT CHANGE THIS
\usepackage{times}  % DO NOT CHANGE THIS
\usepackage{helvet}  % DO NOT CHANGE THIS
\usepackage{courier}  % DO NOT CHANGE THIS
\usepackage[hyphens]{url}  % DO NOT CHANGE THIS
\usepackage{graphicx} % DO NOT CHANGE THIS
\usepackage{natbib}  % DO NOT CHANGE THIS AND DO NOT ADD ANY OPTIONS TO IT
\usepackage{caption} % DO NOT CHANGE THIS AND DO NOT ADD ANY OPTIONS TO IT
\usepackage{algorithm}
\usepackage{algorithmic}

\usepackage{newfloat}
\usepackage{listings}
\DeclareCaptionStyle{ruled}{labelfont=normalfont,labelsep=colon,strut=off} % DO NOT CHANGE THIS
\floatstyle{ruled}
\newfloat{listing}{tb}{lst}{}
\floatname{listing}{Listing}
\title{Ethical Artificial Intelligence Principles and Guidelines for the Governance and Utilization of Highly Advanced Large Language Models}
\author{
    Soaad Qahhār Hossain\textsuperscript{\rm 12},
    Syed Ishtiaque Ahmed\textsuperscript{\rm 13}
}
\affiliations{
    \textsuperscript{\rm 1}Department of Computer Science, University of Toronto\\
    40 St. George Street\\
    Toronto, Ontario, Canada\\
     \textsuperscript{\rm 2}soaad.hossain@mail.utoronto.ca \\
      \textsuperscript{\rm 3}ishtiaque@cs.toronto.edu
}

\usepackage{bibentry}
\begin{document}

\maketitle

\begin{abstract}
Given the success of ChatGPT, LaMDA and other large language models (LLMs), there has been an increase in development and usage of LLMs within the technology sector and other sectors. While the level in which LLMs has not reached a level where it has surpassed human intelligence, there will be a time when it will. Such LLMs can be referred to as advanced LLMs. Currently, there are limited usage of ethical artificial intelligence (AI) principles and guidelines addressing advanced LLMs due to the fact that we have not reached that point yet. However, this is a problem as once we do reach that point, we will not be adequately prepared to deal with the aftermath of it in an ethical and optimal way, which will lead to undesired and unexpected consequences. This paper addresses this issue by discussing what ethical AI principles and guidelines can be used to address highly advanced LLMs.
\end{abstract}

\section{Introduction}
\noindent The research, development and usage of large language models (LLMs) continues to rapidly increase due to its success in demonstrating its abilities to perform tasks more efficiently and effectively than humans. There are limitations with the current LLMs, and those models do make mistakes such as provide incorrect answers. For example, ChatGPT failed the Taiwan’s 2022 Family Medicine Board Exam, only answering 52 questions out of 125 correctly (Tzu-Ling et al. 2023). At some point, these issues will be resolved, but its progress and advancements will not stop there. Research and development in LLMs will come to a point where LLMs will surpass human intelligence, which in this paper we refer to those models as highly advanced LLMs, or simply advanced LLMs. While this can lead to good outcomes through making tasks easier to complete, it can also lead to bad outcomes as well. An example of a bad outcome is when LLMs generate convincingly misleading information that can potentially have negative consequences in areas such as public health and politics (Ray 2023). 

To prevent such issues from occurring, we need to commence the discussion on highly advanced LLMs and discuss how we can approach it. Accordingly, this paper directly addresses this by discussing how ethical artificial intelligence (AI) principles and guidelines could be used on advanced LLMs to prevent them from harming people and the environment. The purpose of the paper is not to comprehensively discuss the applications of the principles and guidelines to advanced LLMs, but rather to initiate the discussion on it and provide details on which principles and guidelines could used to address advanced LLMs, and things that need to be considered for them.

\section{Background}
\noindent Prior to addressing advanced LLMs, we will first review ethical AI principles and guidelines, and LLMs in more details. We will also highlight which principles and guidelines will be used specifically for our approach along with factors to consider.

\subsection{Ethical Artificial Intelligence Principles}
\noindent Ethical AI (also known as AI ethics) principles encompasses a wide range of areas within technology, including responsibility and privacy, fairness, explainability, robustness, transparency, environmental sustainability, inclusion, moral agency, value alignment, accountability, trust, and technology misuse (IBM 2023). It is worth noting that while different areas of the world define ethical AI principles differently, the final goal is more or less the same. For example, the Australian government focuses on 8 ethical AI principles (Australian Government 2023), while the Organization for Economic Cooperation and Development (OECD) focus on 5 (OECD 2019). For both, the goal is to ensure AI is safe, secure, trustworthy and reliable. In our case, when approaching advanced LLMs with ethical AI principles, we will specifically use responsibility, robustness and technology misuse. Future work should investigate the use of other ethical AI principles for advanced LLMs. 

For our approach, we will use the Australian Government's understanding of the ethical AI principle of responsibility. They state that businesses, organisations’ and individuals’ must take responsibility for the outcomes of the AI systems that they design, develop, deploy and operate (Australian Government 2023). For the principle of robustness, we will use a perspective of robustness in AI shared amongst African communities. Their perspective on the principle of robustness states that ethical, legal and socio-cultural impacts of AI need to be robustly considered and mitigated (Eke et al. 2023). We define technology misuse as the intentional use of technology to achieve harmful outcomes (Brundage et al. 2018). Accordingly, the ethical AI principle on technology misuse is to use technology in a way that does not directly or indirectly harm others. 

\subsection{Ethical Artificial Intelligence Guidelines}
\noindent There are multiple organizations that have created their own set of ethical AI guidelines. Such entities include United Nations Educational, Scientific and Cultural Organization (UNESCO), and the EU. UNESCO's guidelines consist of 11 different areas (UNESCO 2022), while the EU's consist of 7. Similar to the ethical AI principles, the final goal for the guidelines is more or less the same. The goal is to protect humanity, individuals, societies and the environment from harms that can be caused from using AI systems. Accordingly, the guidelines that we will use are those concerning societal and environmental well-being, safety, and accountability. Future work should investigate the use of other ethical AI guidelines for advanced LLMs. 

For our approach, the ethical AI guidelines for societal and environmental well-being encompasses health, social well-being, environments and ecosystems. Accordingly, we use the guidelines provided by UNESCO pertaining to those. These guidelines state that organisations' and individuals' designing, developing, deploying and operating should assess the direct and  indirect  environmental  impact  throughout  the  AI  system  life  cycle; that the development and deployment of  AI  systems  related  to  health and mental health is regulated to the effect that they are safe, effective, efficient, scientifically and medically proven and   enable   evidence-based   innovation   and   medical   progress; and more (UNESCO 2022). For the guidelines for safety, we consider it as stated by the EU. The EU states that AI systems need to be resilient and safe, ensuring a fall back plan in case something goes wrong, as well as being accurate, reliable and reproducible (EU 2019). For the guidelines concerning accountability, we will use the guidelines from UNESCO. Their guidelines state that AI systems should be audible and traceable, and that there should be oversight, impact assessment, audit and due diligence mechanisms in place to avoid conflicts with human rights norms and threats to environmental well-being (UNESCO 2022). 

\subsection{Large Language Models and Considerations}
\noindent LLMs are machine learning models trained on massive amounts of text data and are able to generate human-like text, answer questions, and complete other language-related tasks with high accuracy (Ray 2023). It is important to note that current LLMs are experiencing challenges, such as providing incorrect information. An example of this is the Bard disaster case. The Bard disaster case was a case where Bard (the LLM created by Google) provided incorrect information in a demonstration failure that costed Google over 100 billion dollars in stock losses (Floridi 2023). LLMs are currently not at a point where they are not error-free, but they still perform well in general. Normally, when considering LLMs for usage we need to consider legal aspects (such as licensing and commercial use), factors for inference speed and precision, model size, context length, usage type (task specific vs. general purpose), testing and evaluation, deployment, and costs. For our case of advanced LLMs, we will assume that all those considerations are not a problem, and that the usage type considers both task specific and general purpose. That is, there are no limits to the context length, conditions of factors for inference speed and precision are acceptable, etc. Furthermore, we assume that the costs of designing, developing, deploying and operating LLMs to the environment is low.

\section{Comprehending Advanced Large Language Models and its Capabilities}
\noindent To properly approach highly advanced LLMs, we need to understand what they truly are. With the assumptions that we made in the previous section about advanced LLMs and assumption that advanced LLMs have surpassed human intelligence, that would imply several things, which can be encompassed into four implications. First, it would imply that natural language understanding (NLU) including sentiment analysis, text classification, natural language inference and other NLU tasks, reasoning, natural language generation (NLG) including summarization, dialogue, translation, question answering and other NLG tasks, and multilingual tasks are not an issue for advanced LLMs. That is, advanced LLMs will be able to receive data from any language, process and analyze the data, access and utilize existing information (in diverse languages), evaluate information, and produce results in any specified language with ease. Second, it would imply that advanced LLMs can do tasks that are considered extremely difficult for humans and tasks that were originally considered as impossible by humans. For example, it may be able to prove mathematical concepts that were never proven before. Third, it can create new ideas and concepts. Fourth, the amount of resources needed for advanced LLMs is such that it does not harm the environment. This implies that the cost to design, develop and use them is low to an extent where it does not pose a threat. This also means that there are not as many barriers that prevent people and organizations from developing and using them. 

In theory, the first three implications sound exceptionally good. However, in practice they are not. The reason for this is that just as advanced LLMs have the potential of solving good highly difficult problems, such as determining the cure to cancer of any type, it also has the potential of solving bad highly difficult problems, such as determining and explaining what is needed to hack financial institutions. Consequently, the negative impact of advanced LLMs will be catastrophic. Accordingly, measures need to be in-place to prevent such problems from happening. 

\section{Advanced Large Language Models Governance Using AI Ethics}
\noindent The measures that need to be in-place to prevent advanced LLMs from causing massive problems and creating chaos ultimately needs to be policies that are enforced onto the advanced LLM users and developers. Those measures need to have ethical AI principles and guidelines instilled in them. This section will not comprehensively discuss the applications of ethical AI principles and guidelines to advanced LLMs governance, but rather it will focus on selected principles and guidelines, and connecting them to form policies appropriate for advanced LLMs. We will only focus on three policies, detailing their motivations, and elaborating on their use. 

\subsection{Responsibility, Accountability and Advanced Large Language Models}
\noindent The first policy of interest that concerning accountability and responsibility of advanced LLMs. The motivation for this is that this policy will influence the creation and utilization of advanced LLMs. Appropriate policies will instill pressure needed for people and entities planning on creating and using advanced LLMs to create or use them with caution. To approach such policies, we will use the Australian Government's ethical AI principle of responsibility and UNESCO's ethical AI guideline on accountability. Applying the principle, we get the following: businesses, organisations’ and individuals’ must take responsibility for the outcomes of the advanced LLMs that they design, develop, deploy and operate. Similarly, applying the guidelines, we get the following: advanced LLMs should be audible and traceable, and that there should be oversight, impact assessment, audit and due diligence mechanisms in place to avoid conflicts with human rights norms and threats to environmental well-being. The policy should use that principle, but ensure that it is both the developer and the user that are responsible for the outcomes. This will force the developers to incorporate audibility and traceability, oversight, impact assessment, audit and due diligence mechanisms when producing advanced LLMs. In addition, the policies will force user to take precaution when using advanced LLMs as they will be responsible for the outcome. Note that in the case where someone were to use an advanced LLM that led to problematic outcomes, the developers will still be held responsible even though the user was the one that misused the advanced LLM. This is to further enforce audibility, traceability, etc. onto developers. Different from tools such nuclear weapons, we do not know the full extent of damage that could be produced from advanced LLMs due to the fact that advanced LLMs are more intelligent than use. However, we do know that the consequences can be catastrophic. Accordingly, advanced LLMs need to be dealt in a manner that forces those directly and indirectly involved to take responsibility and precaution. 

\subsection{Technology Misuse, Safety and Advanced Large Language Models}
\noindent The second policy of interest is that concerning safety and technology misuse. The motivation for this policy is ultimately that it will protect people and create safety and other measures needed to prevent harm. This policy should focuses on two things. First, it should focuses on the case where something were to go wrong with using advanced LLMs. Second, it should focuses on cases where advanced LLMs is used wrongfully. For the first part, we can use the ethical AI guideline for safety. Using the ethical AI guideline for safety, we get the following: advanced LLMs need to be resilient and safe, ensuring a fall back plan in case something goes wrong, as well as being accurate, reliable and reproducible. To elaborate, in the case where an undesired outcome was produced from using an advanced LLM, it should be possible to recover quickly from such situation. In the case where it is not possible, it should be possible to stop the situation from getting worse, and make corrections when needed. For the second part, we can use the ethical AI principle for technology misuse. Using the principle of technology misuse, we get the following: use advanced LLMs in a way that does not directly or indirectly harm others. This may be difficult for users as not all users know how advanced LLMs can harm others. It is recommended that a license is created to access and operate advanced LLMs. That way only those knowledgeable on advanced LLMs and understand the legal and other aspects attached to its use are able to access and use them. Accordingly, as advanced LLMs are created, the creators of them are required to have them registered with the government so that advanced LLMs are regulated.   

\subsection{Robustness, Societal and Environmental Well-Being and Advanced Large Language Models}
\noindent The third policy of interest is that concerning robustness and societal and environmental well-being. The motivation for the policy is that it will help ensure that the use of advanced LLMs will not damage the well-being of society and the environment. The reason why we did not include development, deployment, etc. of advanced LLMs is because of the assumptions that we made earlier regarding advanced LLMs. We will first apply the ethical AI principle, then discuss the guidelines. In applying the African perspective of ethical AI principle of robustness, we get the following: ethical, legal and socio-cultural impacts of advanced LLMs need to be robustly considered and mitigated. The aim of the policy should be to have advanced LLMs thoroughly considered and mitigated for their ethical, legal and socio-cultural impact prior to their deployment. 

Different from other AI systems and technologies, there are more aspects that need to be considered for advanced LLMs to properly consider and mitigate its ethical, legal and socio-cultural impact. These aspects can be understood through understanding the capabilities of advanced LLMs. For example, knowing that advanced LLMs can do tasks that were originally considered impossible, we can get a sense of what impact it can have on laws. For instance, after investigating use cases of an advanced LLM and the current laws in-place in a country, it can be discovered that the existing laws do not cover certain types of cases. As such, new laws need to be created should that advanced LLM be released. It is recommended that an a screening and approval process similar to that used for pharmaceutical drugs is used for advanced LLMs. Just as pharmaceutical drugs can create large-scale irreversible damages on society and the environment, advanced LLMs can as well. As such, additional parties should be involved in the process of considering and mitigating the ethical, legal and socio-cultural impacts of them.

For the UNESCO's ethical AI guidelines on societal and environmental well-being, they are very comprehensive. The important thing is that in UNESCO's document titled "Recommendation on the Ethics of Artificial Intelligence", the policy recommendations on health, social well-being, environments and ecosystems are applied to the policy concerning robustness and societal and environmental well-being. In applying the guidelines, what would result would be things such as the following: organisations' and individuals' designing, developing, deploying and operating should assess the direct and indirect environmental impact throughout the AI system life cycle; that the development and deployment of  advanced LLMs related to health and mental health is regulated to the effect that they are safe, effective, efficient, scientifically and medically proven and enable evidence-based innovation and medical progress. Applying the guidelines to the policy will help prevent negative consequences to society and the environment from using advanced LLMs.

\section{Considerations for Advanced Large Language Models and Policy-Making}
\noindent While advanced LLMs are still considered as AI, it needs to be treated differently from other AI systems because unlike other AI systems, advanced LLMs can complete tasks considered highly difficult or impossible by human beings, and formulate ideas and concepts. This differs from other forms of AI as other forms of AI cannot complete tasks impossible to humans and generate new ideas. Accordingly, policies on advanced LLMs should consider both things can be achieved using advanced LLMs given what we know about advanced LLMs and things that not likely to occur or be achieved using advanced LLMs given our present knowledge of LLMs. While certain tasks may not look achievable in the present moment, it does not mean that it will never be achieved. Given that advanced LLMs are more intelligent than us, it is only a matter of time before those tasks are completed.  

\subsection{Considerations for Impact of Policies on Utility}
\noindent An aspect that should be considered when creating policies for advanced LLMs is the impact that the various policies can have on the utility generated by advanced LLMs. While the impact largely depends on the policy itself, in general there are some general negative and positive impacts that can result from the various policies. We will discuss the main negative and positive impact, and selected other impacts. 

The main negative impact that policies will have on the utility generated by advanced LLMs is that it to a certain level, it will prevent quick action and progress from taking place. In a sense this is good as this applies to cyberattacks and other forms of attacks. However, it is bad in the sense that it prevents quick action that can lead to good outcomes from taking place. This does not mean that it will stop such actions from taking place altogether, but it means that it will take more time for such action to take place. For example, certain computations, summaries and analyses may take longer to complete due to some individuals not having or having limited access to advanced LLMs. Consequently, this may delay the speed of progress in achieving scientific and other goals and objectives. Other negative impacts that policies will have on the utility generated by advanced LLMs is that it can limit the use cases of advanced LLMs. With less people using advanced LLM, it will take more time to understand the full extend of advanced LLMs in terms of utility and production. Consequently, the use case of advanced LLMs will be limited for a period of time. Note that this can be addressed through comprehending the use cases for existing LLMs and emerging technologies (e.g. chatbots). 

The easiest way to comprehend the main positive impact that policies will have on the utility generated by advanced LLMs is by looking at the the utility around cyberspace and cyberpower. We define cyberspace and cyberpower like how Colonel Kevin L. Parker, U.S.A. Air Force, defines it. Cyberspace is defined as the domain that exists for inputting, storing, transmitting, and extracting information utilizing the electromagnetic spectrum; and cyberpower is defined as the potential to use cyberspace to achieve desired outcomes (Parker 2014). With cyberspace, it favors offense because it enables quick action and concentration, allows anonymity, and expands the spectrum of nonlethal weapons (Parker 2014). In the same sense, the utility generated by advanced LLMs without policies in-place allow for quick action and concentration, anonymity, and expansion of nonlethal weapons. These motivate and create more opportunities for cyberattacks, which is highly problematic as cyberattacks can result in intended and unintended consequences impacting individuals and populations (Parker 2014). With anonymity and other factors, the problem is further exacerbated as challenges around accountability and responsibility are added to the issues. Ultimately, the utility generated by advanced LLMs without policies will increase the likelihood of cyberattacks. In having policies in-place, they limit the utility generated by advanced LLMs, subsequently decreasing the volume of cyberattacks and likelihood of cyberattacks from occurring. In turn, the amount of intended and unintended consequences from cyberattacks, and other problems associated with those (e.g. social determinants of health, environmental impact), will decrease. Other positive impacts that policies will have on the utility generated by advanced LLMs is that it will force advanced LLMs to be used more purposefully, enabling for important and pressing issues to be prioritized. The limited access, cost of resources, potential punishments, etc. will force users to focus more on needs than wants when using advanced LLMs, and will force users to be less reckless when using advanced LLMs. 

\subsection{Trade-off Consideration Between Consequences and Utility}
\noindent When making policies for advanced LLMs, arguably the biggest component that needs to be considered is the trade-off between the consequences and the utility generated from advanced LLMs. A way of viewing the trade-off is viewing advanced LLMs as a weapon of mass destruction like a thermonuclear bomb. On one side, thermonuclear bombs can be used to protect countries from other countries. On the other, using them will cause catastrophic damage to both the country received and delivering the thermonuclear bomb. The impact to the country being thermonuclear bombed will range from psychological damage to the population to economic damage to the country affected. The impact to the country using the thermonuclear bombs will range from political damage (including international relations, government and entities associated with it, etc.) to economic damage (including talent migration, boycotting, etc.). What is key in such case consequences associated with using the thermonuclear bomb. If there were little to no consequences from the user with using such weapon of mass destruction, then it would be used in wars and conflicts following World War 2, but the reality is that there are consequences to the users and those consequences are severe. 

Advanced LLMs can be as destructive as weapons of mass destruction. In the wrong hands, it can be used to realize extremely harmful things such as highly potent bioweapons, large-scale cyberattacks (e.g. mass security breaches, governmental-level hacking), and psychological warefare. These things will not only harm the intended group its supposed to harm, but will harm the user without them realizing it (e.g. feeling of guilt after using advanced LLMs to create a new variant of a disease), and can hurt those connected to the user. The type of harm the user will receive would ultimately depend on what it was used for, but the impact it will have on the user will result in some sort of unintended consequence that will negatively impact them personally, socially, financially, etc. The way in which those connected to the user can be hurt depends on their relationship with the user. Ultimately there are consequences for those involved in using advanced LLMs, and to some extent consequences for those indirectly involved. These consequences need to be considered when evaluating policies and the utility of advanced LLMs. 

\section{Discussion}
\noindent The existing ethical AI principles and guidelines can be used to approach the utilization of highly advanced LLMs. In addition, policies using specific ethical AI principles and guidelines can address both areas that they were intended to address, and areas that were not intended to be addressed but need to be addressed through policy. For instance, the policy on safety and technology misuse compliments the policy on responsibility and accountability as it elaborates on areas pertaining to the advanced LLM users. While policies on advanced LLMs can and should use existing ethical AI principles and guidelines, they should also consider policies and approaches that we use for matters that are not technical, such as guns and pharmaceutical drugs. Advanced LLMs have the potential of causing catastrophic damage to society and the environment. Accordingly, we need to carefully approach it like how we approached other things that can cause significant harm. We also need to make sure that we do not create more problems in attempts to solve selected existing problems when using advanced LLMs.

\section{Conclusion}
\noindent Advanced LLMs policies need to be created prior to prevent advanced LLMs from causing hard to society and the environment. Those policies can be created using existing ethical AI principles and guidelines. However, the policies should not limit itself to AI as highly advanced LLMs will be capable of doing things that other AI systems are not able to do. Future work should investigate the use of other ethical AI principles and guidelines for policy-making for advanced LLMs, the full potential of highly advanced LLMs, and the impact of policies on highly advanced LLMs.

\nocite{*}
\bibliography{aaai24}

\end{document}